\documentclass[PRL,twocolumn,showpacs,preprintnumbers,amsmath,amssymb,nofootinbib]{revtex4-2}
\usepackage{graphicx}
\usepackage{dcolumn}
\usepackage{bm}
\usepackage{hyperref}
\usepackage{orcidlink}
\usepackage{dsfont}

\begin{document}

\title{Separating entanglement at emission from femtoscopic correlations}

\author{Isaac Vida\~na\orcidlink{0000-0001-7930-9112}}
\affiliation{Istituto Nazionale di Fisica Nucleare, Sezione di Catania, Dipartimento di Fisica ``Ettore Majorana'', Universit\`a di Catania, Via Santa Sofia 64, I-95123 Catania, Italy}

\date{\today}


\begin{abstract}
We investigate how femtoscopic correlations
modify the relation between the spin
entanglement of emitted particle pairs and
that of pairs selected at a given relative
momentum. For two spin-$\frac{1}{2}$ particles,
we show that the momentum-selected spin
state can be entangled even when the
corresponding state at emission is separable.
We derive a momentum-dependent separability
criterion that determines whether the
observed spin correlations require
entanglement at emission.
As an application, we consider a pair of
identical $\Lambda$ hyperons, for which
quantum statistics and final-state interactions
affect the singlet and triplet spin channels
differently at low relative momentum.
Our results provide a framework for using
experimentally accessible spin correlations
to investigate entanglement at emission.
\end{abstract}

\maketitle



Quantum entanglement has become an increasingly relevant subject in particle and nuclear physics, where correlations between the internal degrees of freedom of produced particles offer the possibility of investigating the quantum nature of particle production processes~\cite{Gong2022,Barata2024}.
Pairs of spin-$\frac{1}{2}$ particles are
particularly suitable for this purpose,
since their spin correlations can provide
information on the corresponding two-particle
quantum state and its entanglement 
properties~\cite{Gong2022,Lednicky2001}.
Particular care is required, however, when interpreting spin correlations of particle pairs with low relative momentum. In this region, quantum statistics and final-state interactions can influence the 
spin-singlet and spin-triplet components of the two-particle state differently. As a result, the relative populations of these components in the observed pairs may differ from those at emission.

Femtoscopy provides a method to investigate the space-time characteristics of particle-emitting sources and the interactions among particles produced in high-energy collisions through momentum correlations~\cite{STAR2013,ALICE2015,ALICE2019,Fabbietti2021}. These correlations arise from quantum statistics and final-state interactions, and may depend on the spin configuration of the particles involved. For pairs of spin-$\frac{1}{2}$ particles, this dependence can be expressed in terms of the spin-singlet and spin-triplet components. Recently, Zhang {\it et al.}~\cite{Zhang2026}  
proposed a spin-femtoscopy framework to determine the relative populations of these spin states from spin-dependent femtoscopic correlation functions. Fan {\it et al.}~\cite{Fan2026} subsequently
showed that femtoscopic effects can give rise
to spin correlations in the observed pairs
even when the spins of the emitted particles
are initially uncorrelated. In the present work, we show that the spin
state of particle pairs selected at a given
relative momentum can be entangled even when
the corresponding state at emission is
separable, and derive a criterion for
identifying entanglement at emission from
the observed spin correlations.


We first consider the spin state of a particle pair at emission
and examine how femtoscopic correlations modify its spin
composition. We restrict our analysis to rotationally
invariant two-particle spin states, an
approximation appropriate when single-particle
polarizations and anisotropies in the spin
correlations can be neglected. Under this assumption, the spin state of a pair of
spin-$\frac{1}{2}$ particles can be described as a mixture
of a spin-singlet state and three equally populated spin-triplet
states. Allowing the spin composition at emission to depend
on the relative momentum $q$, the corresponding two-body
spin density matrix can be written as
\begin{equation}
\rho^{\rm emi}(q)
=
w_0^{\rm emi}(q)|00\rangle\langle 00|
+
\frac{1-w_0^{\rm emi}(q)}{3}
\sum_{m=-1}^{1}|1m\rangle\langle 1m|\ ,
\label{eq:rho_emi}
\end{equation}
where $|Sm\rangle$ denotes a two-particle state with total
spin $S$ and spin projection $m$, and the label ``emi''
refers to quantities evaluated at emission. The quantity $w_0^{\rm emi}(q)$ represents the singlet fraction of the emitted pairs at relative momentum $q$, whereas
$1-w_0^{\rm emi}(q)$ is the
corresponding triplet fraction. In
particular, for a pair of particles emitted with unpolarized
and uncorrelated spins, the four spin states are equally
populated, giving $w_0^{\rm emi}(q)=1/4$.
The factor $1/3$ ensures equal populations
of the three triplet substates, as required by rotational invariance.

We now examine how femtoscopic correlations modify the
spin composition of the emitted pair. In the following,
we consider spin-conserving interactions, for which
the singlet and triplet channels can be treated
independently. For the rotationally
invariant spin states considered here, the effects of
quantum statistics and final-state interactions on the pair yields can be
characterized by two spin-dependent femtoscopic correlation
functions, $C_0(q)$ and $C_1(q)$, associated with the
spin-singlet and spin-triplet channels, respectively.
Here $q=|\mathbf{p}_1-\mathbf{p}_2|/2$ denotes the
magnitude of the relative momentum in the pair rest frame.

The functions $C_0(q)$ and $C_1(q)$ describe the relative
enhancement or suppression of spin-singlet and spin-triplet
pairs at a given relative momentum $q$ with respect to an
uncorrelated reference. Values of $C_S(q)>1$ ($C_S(q)<1$)
indicate an enhancement (suppression) of the pair yield,
which may arise from attractive or repulsive final-state interactions, as well as from quantum-statistical effects associated with the symmetry of the two-particle wave function.

We now consider how the different femtoscopic correlations
in the singlet and triplet channels affect the spin
composition of the observed pairs. 
Let
$N_S^{\rm emi}(q)$ and $N_S^{\rm obs}(q)$ denote
the pair yields in a narrow interval around the
relative momentum $q$, for total spin $S=0,1$,
at emission and after femtoscopic correlations
are included, respectively.

The singlet fraction at emission is then
\begin{equation}
w_0^{\rm emi}(q)
=
\frac{N_0^{\rm emi}(q)}
{N_0^{\rm emi}(q)+N_1^{\rm emi}(q)} \ .
\label{eq:w_emi}
\end{equation}
The spin-dependent femtoscopic correlation functions
relate the observed pair yields to those at emission
according to
\begin{equation}
N_S^{\rm obs}(q)
=
C_S(q)N_S^{\rm emi}(q) \ ,
\qquad S=0,1 \ .
\label{eq:yields}
\end{equation}
The singlet fraction of the observed pairs is
obtained by dividing the observed singlet yield
by the total observed pair yield. Using
Eq.~(\ref{eq:yields}), we obtain
\begin{equation}
w_0^{\rm obs}(q)
=
\frac{N_0^{\rm emi}(q)C_0(q)}
{N_0^{\rm emi}(q)C_0(q)
+N_1^{\rm emi}(q)C_1(q)} \ .
\label{eq:w_obs_intermediate}
\end{equation}
Dividing numerator and denominator by the total
pair yield at emission,
$N_0^{\rm emi}(q)+N_1^{\rm emi}(q)$,
and using Eq.~(\ref{eq:w_emi}), we finally obtain
\begin{equation}
w_0^{\rm obs}(q)
=
\frac{w_0^{\rm emi}(q)C_0(q)}
{w_0^{\rm emi}(q)C_0(q)
+[1-w_0^{\rm emi}(q)]C_1(q)} \ .
\label{eq:w_obs}
\end{equation}
Equation~(\ref{eq:w_obs}) shows that the singlet
fraction of the observed pairs can differ
from that at emission when the singlet and
triplet femtoscopic correlation functions are
different.

Having established how femtoscopic correlations modify
the singlet fraction of the observed pairs, we now
examine the consequences for their entanglement
properties. For this purpose, we introduce the
spin-correlation parameter $P_{\rm obs}(q)$,
which characterizes the average correlation
between the spins of the two particles at
relative momentum $q$. It is defined as \cite{Lednicky2001}
\begin{equation}
P_{\rm obs}(q)
=
\frac{1}{3}
\left\langle
\boldsymbol{\sigma}_1\cdot
\boldsymbol{\sigma}_2
\right\rangle_q
=\frac{1}{3}\mathrm{Tr}\left[
\rho^{\rm obs}(q)\,
\boldsymbol{\sigma}_1\cdot\boldsymbol{\sigma}_2
\right] \ .
\label{eq:Pdef}
\end{equation}
where the spin density matrix $\rho^{\rm obs}(q)$ of the observed pair has the same form as Eq.~(\ref{eq:rho_emi}), with the singlet fraction of emitted pairs, $w_0^{\rm emi}(q)$, replaced by that of the observed ones, $w_0^{\rm obs}(q)$.

Since the operator
$\boldsymbol{\sigma}_1\cdot
\boldsymbol{\sigma}_2$
has eigenvalue $-3$ in the singlet state
and $+1$ in the triplet states, we obtain
\begin{equation}
P_{\rm obs}(q)
=
\frac{1-4w_0^{\rm obs}(q)}{3}\ .
\label{eq:P_w}
\end{equation}

The entanglement properties of the two-body spin state
can be established using the Peres--Horodecki
separability criterion, also known as the positive
partial transpose (PPT) criterion~\cite{Peres1996,Horodecki1996}.
For a pair of spin-$\frac{1}{2}$ particles, this
criterion states that their two-body spin density
matrix is separable if and only if all eigenvalues
of its partial transpose\footnote{The partial
transpose $\rho^{T_2}$ of a density matrix $\rho$
is obtained by exchanging the bra and ket indices
of the second particle while leaving those of
the first unchanged, {\it i.e.},
$\langle ij|\rho^{T_2}|kl\rangle
=\langle il|\rho|kj\rangle$.}
are nonnegative.

For the rotationally invariant density matrix
$\rho^{\rm obs}(q)$ considered here, the eigenvalues
of its partial transpose are
\begin{equation}
\lambda_{1,2,3}(q)
=
\frac{1+2w_0^{\rm obs}(q)}{6},
\qquad
\lambda_4(q)
=
\frac{1}{2}-w_0^{\rm obs}(q) \ .
\label{eq:ppt_eigenvalues}
\end{equation}
Since the first three eigenvalues are always
nonnegative, the separability condition is
determined by the fourth one.
Consequently, the observed spin state is entangled when
\begin{equation}
\lambda_4(q)<0
\quad\Longleftrightarrow\quad
w_0^{\rm obs}(q)>\frac{1}{2} \ .
\label{eq:ent_obs_w}
\end{equation}
or, equivalently, using Eq.~(\ref{eq:P_w}),
\begin{equation}
P_{\rm obs}(q)<-\frac{1}{3} \ .
\label{eq:ent_obs_P}
\end{equation}

Equation~(\ref{eq:ent_obs_w}) (or (\ref{eq:ent_obs_P})) establishes the
entanglement condition of the observed pairs, but does not
necessarily imply that their spin state was already
entangled at emission. Indeed, as shown above,
femtoscopic correlations can modify the singlet
fraction between emission and observation. 
To determine whether the emitted state was
entangled, we apply the same PPT criterion to
the density matrix $\rho^{\rm emi}(q)$. Since
this matrix has the same rotationally invariant
form as $\rho^{\rm obs}(q)$, entanglement at
emission requires
\begin{equation}
w_0^{\rm emi}(q)>\frac{1}{2} \ .
\label{eq:ent_emi_w}
\end{equation}

To express this condition in terms of the observed
spin correlations, we introduce the critical value
$P_{\rm crit}(q)$, defined as the value of
$P_{\rm obs}(q)$ corresponding to the separability
boundary $w_0^{\rm emi}(q)=1/2$.
Using Eqs.~(\ref{eq:w_obs}) and (\ref{eq:P_w}),
we obtain
\begin{equation}
P_{\rm crit}(q)
=
-\frac{3C_0(q)-C_1(q)}
{3[C_0(q)+C_1(q)]}.
\label{eq:P_crit}
\end{equation}
Consequently, the emitted state is entangled
if and only if\footnote{This condition requires both $C_0(q)$
and $C_1(q)$ to be nonzero. If either vanishes,
the corresponding spin channel is completely
suppressed, so that the observed pairs belong
entirely to the other channel, regardless of
its initial population. Consequently, the
observed spin state cannot determine whether
the emitted state was entangled.}
\begin{equation}
P_{\rm obs}(q)<P_{\rm crit}(q) \ .
\label{eq:ent_emi_P}
\end{equation}
This condition provides a momentum-dependent
criterion for identifying entanglement at emission
from the spin correlations of the observed pairs,
accounting for the different femtoscopic responses
of the singlet and triplet channels. We note that when $C_0(q)=C_1(q)\neq 0$,
the singlet fractions at emission and observation
coincide [see Eq.~(\ref{eq:w_obs})], and
$P_{\rm crit}(q)=-1/3$. Consequently, the
entanglement conditions for the emitted and
observed pairs become identical, as given by
Eq.~(\ref{eq:ent_obs_P}).

An important consequence of this result is that
the spin state of particle pairs observed at a
given relative momentum can be entangled even
when the corresponding state at emission is
separable. This is possible when the singlet
channel is sufficiently enhanced relative to
the triplet channel, so that the observed
singlet fraction can exceed $1/2$ even if
$w_0^{\rm emi}(q)\leq 1/2$.
In this case, the observed spin correlations
can satisfy $P_{\rm obs}(q)<-1/3$ without
satisfying $P_{\rm obs}(q)<P_{\rm crit}(q)$,
showing that entanglement of the observed
pairs does not necessarily imply entanglement
at emission.

To illustrate an application of our separability
criterion, we consider a pair of identical
$\Lambda$ hyperons and evaluate the spin-dependent
femtoscopic correlation functions entering
$P_{\rm crit}(q)$.

The spin-dependent femtoscopic correlation functions
are calculated using the Koonin--Pratt formalism
\cite{Koonin1977,Pratt1986}, in which
\begin{equation}
C_S(q)=\int d^3r\,S(r)
\left|\Psi_{qS}^{(-)}(\mathbf r)\right|^2,
\label{eq:KP}
\end{equation}
where $S(r)$ is the normalized distribution of
relative emission distances and
$\Psi_{qS}^{(-)}(\mathbf r)$ is the relative
two-particle scattering wave function in the
spin channel $S=0,1$, including quantum
statistics and final-state interactions.
We assume a spherically symmetric Gaussian
source,
\begin{equation}
S(r)=\frac{1}{(4\pi R^2)^{3/2}}
\exp\left(-\frac{r^2}{4R^2}\right),
\label{eq:source}
\end{equation}
where $R$ characterizes the spatial extent
of the emitting source.

For two identical $\Lambda$ hyperons, the total
isospin is $I=0$, and the antisymmetry of the
two-fermion wave function requires $L+S$
to be even, where $L$ and $S$ denote the
orbital angular momentum and total spin,
respectively. Consequently, only partial waves with even (odd) orbital angular momentum $L$ are allowed in the spin-singlet (spin-triplet) channel. At low relative momentum, the $S$-wave
($L=0$) dominates. The ${}^{1}S_0$ channel
is therefore the leading contribution,
while the ${}^{3}S_1$ channel is forbidden
by the Pauli principle.

In the absence of final-state interactions,
the exchange symmetry of the relative wave function
gives
\begin{equation}
C_0^{\rm QS}(q)=1+e^{-4q^2R^2},
\qquad
C_1^{\rm QS}(q)=1-e^{-4q^2R^2},
\label{eq:C_QS}
\end{equation}
where QS denotes quantum statistics.
Thus, at low relative momentum, quantum
statistics enhances the singlet correlation
and suppresses the triplet one.

The strong final-state interaction is included
using the Lednick\'y--Lyuboshitz approximation
\cite{LednickyLL}, retaining the dominant
${}^{1}S_0$ contribution at low relative
momentum. The resulting singlet correlation
function is
\begin{widetext}
\begin{equation}
\begin{aligned}
C_0(q)=1+e^{-4q^2R^2}
&+\frac{|f(q)|^2}{R^2}
\left(1-\frac{d_0}
{2\sqrt{\pi}R}\right)
+\frac{4\,\mathrm{Re}\,f(q)}
{\sqrt{\pi}R}F_1(2qR)
-\frac{2\,\mathrm{Im}\,f(q)}{R}
F_2(2qR),
\end{aligned}
\label{eq:C0_LL}
\end{equation}
\end{widetext}
where $f(q)$ is the ${}^{1}S_0$ scattering
amplitude, parametrized through the
effective-range expansion as
\begin{equation}
f(q)=
\left[
\frac{1}{f_0}
+\frac{d_0}{2}q^2-iq
\right]^{-1}.
\label{eq:f_scatt}
\end{equation}
Here $f_0$ and $d_0$ are the scattering
length and effective range, respectively, with
the scattering-length sign convention adopted by the ALICE Collaboration~\cite{ALICE2019}. The functions $F_1$ and $F_2$ are defined by
$F_1(z)=z^{-1}\int_0^z e^{x^2-z^2}\,dx$
and $F_2(z)=(1-e^{-z^2})/z$.

Neglecting higher-partial-wave final-state
interactions, the triplet correlation function
remains $C_1(q)=C_1^{\rm QS}(q)$.

We evaluate the spin-dependent femtoscopic
correlation functions within the
Lednick\'y--Lyuboshitz approximation using
the $\Lambda\Lambda$ scattering parameters
obtained from the NLO16 interaction developed
by Haidenbauer {\it et al.}~\cite{Haidenbauer2016}.
In particular, we use the values $f_0=0.66$ fm
and $d_0=5.05$ fm corresponding to a regulator
cutoff of $600$ MeV.
We have changed the sign of the scattering length
with respect to Ref.~\cite{Haidenbauer2016}
to be consistent with the ALICE convention
adopted here. We consider Gaussian source radii $R=1.2$
and $1.4$ fm, representative of those
used in ALICE $\Lambda\Lambda$ femtoscopic
analyses~\cite{ALICELL2019}. The reliability of the Lednick\'y--Lyuboshitz
approximation at low relative momentum is
supported by its agreement with a full
Koonin--Pratt calculation using the NLO16
interaction for $R=1.2$ fm~\cite{Haidenbauer2019}.

\begin{figure}[t]
\centering
\includegraphics[width=\columnwidth]
{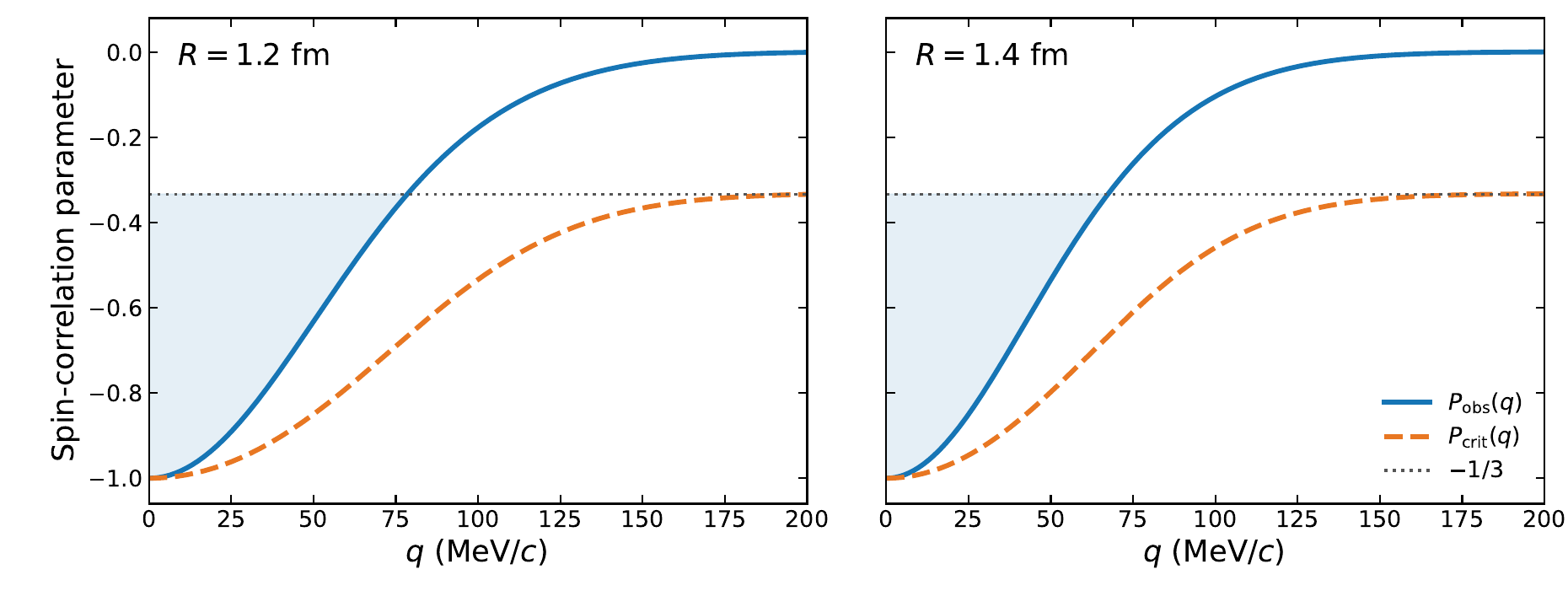}
\caption{Relative-momentum dependence of the
observed spin-correlation parameter
$P_{\rm obs}(q)$ (solid lines) and the critical
value $P_{\rm crit}(q)$ (dashed lines) for
$\Lambda\Lambda$ pairs with initially unpolarized
and uncorrelated spins, $w_0^{\rm emi}=1/4$.
Results are shown for Gaussian source radii
$R=1.2$ and $1.4$ fm, using the NLO16
$\Lambda\Lambda$ scattering parameters within
the Lednick\'y--Lyuboshitz approximation.
The horizontal dotted line indicates the
entanglement threshold $P_{\rm obs}=-1/3$.
Below this threshold, the observed pairs are
entangled, although their spin state at emission
is separable.}
\label{fig:entanglement}
\end{figure}

To illustrate how femtoscopic correlations
can affect the entanglement properties of
the observed pairs, we consider an initially
unpolarized and uncorrelated spin state,
corresponding to $w_0^{\rm emi}(q)=1/4$.
The state at emission is therefore separable.
Figure~\ref{fig:entanglement} shows the resulting
spin-correlation parameter $P_{\rm obs}(q)$
together with the critical value $P_{\rm crit}(q)$
for the two source radii considered. 
As shown in the figure,
$P_{\rm obs}(q)$ falls below the entanglement
threshold $-1/3$ at low relative momentum,
indicating that the observed pairs are entangled
although their spin state at emission is
separable. This occurs for $q$ below approximately
$78$ MeV/$c$ for $R=1.2$ fm and $67$ MeV/$c$
for $R=1.4$ fm. In the limit $q\to 0$, the triplet channel
is completely suppressed and the observed
spin state approaches a pure singlet,
$P_{\rm obs}(q)\to -1$, regardless of whether
the emitted state was separable or entangled.
We note that $P_{\rm obs}(q)$ remains above
$P_{\rm crit}(q)$ for $q>0$, consistently
indicating that the state at emission is
separable. The momentum range over which
the observed pairs are entangled becomes
narrower for the larger source radius.

The spin-correlation parameter $P_{\rm obs}(q)$
can be determined experimentally in bins of
relative momentum $q$ from the parity-violating
weak decays $\Lambda\to p\pi^-$ of the two
hyperons~\cite{Lednicky2001,STARSpin2026}. The angular correlation between
the decay protons is given
by
\begin{equation}
\frac{1}{N}\frac{dN}{d\cos\theta}
=
\frac{1}{2}
\left[1+\alpha_\Lambda^2
P_{\rm obs}(q)\cos\theta\right],
\label{eq:decay_corr}
\end{equation}
where $\theta$ is the angle between the
directions of the two decay protons, each
measured in the rest frame of its parent
$\Lambda$ with respect to consistently
oriented axes, and $\alpha_\Lambda$ is
the weak-decay asymmetry parameter,
which characterizes the correlation between
the $\Lambda$ spin and the direction of
the emitted proton. The measured $P_{\rm obs}(q)$ can then be compared with $P_{\rm crit}(q)$ to determine whether the observed spin correlations
require entanglement at emission.

In summary, we have shown that femtoscopic
correlations can make the spin state of
momentum-selected particle pairs entangled
even when their state at emission is separable.
By combining the Peres--Horodecki separability
criterion with spin-dependent femtoscopic
correlation functions, we have derived a
momentum-dependent condition that distinguishes
entanglement at emission from that of the
observed pairs. This provides a framework
for interpreting measured spin correlations
and assessing whether they constitute
evidence of entanglement at emission.

The author is very grateful to Dieterich Unkel for the useful comments and discussions held during the course of this work.



\end{document}